\documentclass[preprint]{aastex631}
\shorttitle{Li Depletion and Li Dip}
\shortauthors{Li et al.}
\usepackage{hyperref}
\usepackage{placeins}
\usepackage{changepage}
\graphicspath{{./}{figures/}}

\begin{document}
\renewcommand{\thefootnote}{\roman{footnote}}
\title{Meridional Circulation II: A Unified Mechanism for Lithium Depletion in Solar Analogs and the Lithium Dip in Mid-F Cluster Stars}

\email{lixuefeng@ynao.ac.cn}

\author[0000-0001-9291-4261]{Xue-Feng Li}
\affiliation{Yunnan Observatories, Chinese Academy of Sciences, P.O. Box110, Kunming 650216, China}

\author[0000-0002-0349-7839]{Jian-Rong Shi}
\affiliation{CAS Key Laboratory of Optical Astronomy, National Astronomical Observatories, Beijing 100101, China}
\affiliation{University of Chinese Academy of Sciences, Beijing 100049, China}

\author[0000-0002-1424-3164]{Yan Li}
\affiliation{Yunnan Observatories, Chinese Academy of Sciences, P.O. Box110, Kunming 650216, China}
\affiliation{University of Chinese Academy of Sciences, Beijing 100049, China}
\affiliation{Key Laboratory for Structure and Evolution of Celestial Objects, Chinese   Academy of Sciences, P.O. Box110, Kunming 650216, China}
\affiliation{Center for Astronomical Mega-Science, Chinese Academy of Sciences, Beijing 100012, China}

\author[0000-0002-8609-3599]{Hong-Liang Yan}
\affiliation{CAS Key Laboratory of Optical Astronomy, National Astronomical Observatories, Beijing 100101, China}
\affiliation{University of Chinese Academy of Sciences, Beijing 100049, China}
\affiliation{Institute for Frontiers in Astronomy and Astrophysics, Beijing Normal University,  Beijing 102206, China}

\author[0000-0002-2510-6931]{Jing-Hua Zhang}
\affiliation{South-Western Institute for Astronomy Research, Yunnan University, Chenggong District, Kunming 650500, China}



\begin{abstract}
The behavior of lithium (Li) in Population I main sequence stars challenges standard stellar theory. Two phenomena stand out: the solar Li problem which extends to Li depletion in solar analogs and the Li dip observed in mid-F stars within open clusters. Building on the meridional circulation-driven radial mixing framework previously developed to explain Li-enriched red clump stars, we explore its relevance to Li depletion on the main sequence. First, our models reproduce the observed $A(\text{Li})$-Age correlation in solar analogs. Through detailed isochrone analysis, we find good agreement between the simulated and observed $A(\text{Li})$-$T_{\text{eff}}$ relationships within the solar analog parameter space. However, the predicted solar Li abundance ($\sim 1.5\,\text{dex}$) is still higher than current solar measurements. Second, our models partially explain the Li dip phenomenon in mid-F cluster stars. The models accurately reproduce Li distributions on the cool side of the Li dip in most clusters and capture the Li behaviors on the hot side observed in systems like the Hyades. However, we identify limitations in the models' ability to fully reproduce the dip morphology, particularly due to the rotation velocity distribution of sample stars in this temperature zone.

\end{abstract}



\section{Introduction}\label{sect1}

As observational capabilities advance, our understanding of lithium (Li) in the universe has grown clearer. However, significant discrepancies persist between theoretical predictions for Li and observations, particularly three stellar Li problems: 
(1) G/K dwarfs exhibit notable Li depletion, a phenomenon exemplified by the solar Li problem. In this case, observations indicate a $A(\rm Li)$\footnote{Li abundance: $A(\rm Li)=log$$[N(\text{Li})/N(\text{H})]+12$.} of approximately $1.0\,\rm dex$ \citep{1997AJ....113.1871K}, far below the model-predicted value of $2.8\,\rm dex$ \citep{1997ARA&A..35..557P}; 
(2) The distribution of Li for open cluster mid-F stars shows a dip (gap) behavior \citep{1986ApJ...302L..49B,1993AJ....106.1080S}; and 
(3) the `1\% puzzle' of Li-rich, $A(\rm Li)>1.5\,dex$, giants \citep{2019ApJS..245...33G,2021MNRAS.505.5340M,2024MNRAS.529.1423L}. 
Additionally, Li in the Galaxy faces unresolved nucleosynthetic origins \citep{2018A&A...610A..38F,2019MNRAS.482.4372C,2024A&A...691A.142B}, while the cosmological Li problem reveals a $3-4$ times discrepancy \citep{2021A&ARv..29....5M}. Based on baryon density measurements and nuclear reaction rate determinations \citep{1985ARA&A..23..319B}, the predicted value of Big Bang nucleosynthesis \citep[$\sim 2.7\,\rm dex$;][]{2008JCAP...11..012C,2011ARNPS..61...47F,2014PhRvL.113d2501A} exceeds the observed Li plateau in metal-poor stars \citep[$\sim 2.2\,\rm dex$;][]{1982A&A...115..357S,2010A&A...522A..26S,2021MNRAS.505..200M}.

The current solar Li abundance is approximately $2 \,\rm dex$ lower than that of meteorites \citep{1989GeCoA..53..197A,1998SSRv...85..161G,2009ARA&A..47..481A}. Studies of older solar analogs reveal even more significant Li depletion \citep[e.g.][]{2016A&A...587A.100C, 2024arXiv241017590Y}. This indicates a clear $A(\rm Li)$ and age correlation in solar analogs \citep{2019MNRAS.485.4052C}. According to the standard solar model, such pronounced Li depletion should not occur \citep[e.g.][]{1957ApJ...125..233S}, implying that non-standard processes play a crucial role in Li depletion. Many physical processes come into view, including convective overshooting \citep{1999A&A...347..272S, 2017ApJ...845L...6B}, rotation mixing \citep{1981ApJ...243..625E, 1992A&A...255..191C, 2005Sci...309.2189C, 2016ApJ...829...32S}, mixing driven by internal gravity waves \citep{1991ApJ...377..268G, 1994A&A...281..421M}, element diffusion \citep{1986ApJ...302..650M, 1993ApJ...416..312R, 1998ApJ...504..559T}, turbulent diffusion \citep{1985A&A...149..309B}, tachocline diffusion \citep{2003A&A...408.1037P}, mass loss effects \citep{1992ApJ...395..654S}, convective settling \citep{2015A&A...579A.122A}, macroscopic mixing \citep{2025arXiv250103723B}, and the planet effects \citep{2002A&A...386.1039M, 2009A&A...494..663C}. 
Despite the numerous physical processes that can cause Li depletion, there is still no consensus on the dominant mechanism. More research is needed to fully understand how Li depletion occurs.

The Li dip was first discovered in the Hyades \citep{1986ApJ...302L..49B}. This dip primarily locates the effective temperature ($T_{\rm eff}$) range of 7000–6200 K (corresponding to mid-F dwarfs), where the Li abundance significantly declines by approximately 2–3 dex. Subsequently, similar Li distribution features have been identified in numerous clusters, such as NGC 752 \citep{1986ApJ...309L..17H}, Pleiades \citep{1988ApJ...327..389B}, Praesepe \citep{1993AJ....106.1080S}, M67 \citep{1995ApJ...446..203B}, NGC 3680 \citep{2009AJ....138.1171A}, NGC 2243 \citep{2021AJ....161..159A}, NGC 3532, NGC 2420, Ruprecht 134 \citep{2021FrASS...8....6R}, and M48 \citep{2023ApJ...952...71S}. \citet{1988ApJ...334..746C} found that meridional circulation could explain the cold side of the Li dip based on the model proposed by \citet{1982ApJS...49..317T}, while the high Li abundance feature on the hot side may be dominated by radiative acceleration. Additionally, the diffusion of elements within non-rotating stars can replicate the Li dip behavior, but it is narrower than that observed in these clusters \citep[see][]{1993ApJ...416..312R}. Moreover, as anticipated by \citet{1988ApJ...334..746C}, element diffusion leads to Li overabundance on the hot side of the Li dip. Furthermore, mass loss and wave-driven mixing are also possible mechanisms \citep{1990ApJ...359L..55S, 1996A&A...305..513M}. In general, these physical mechanisms are too weak to account for the hot side of the Li dip.

The classical description of meridional circulation can be summarized as the Eddington–Sweet circulation model \citep{1925Obs....48...73E, 1950MNRAS.110..548S}, which proposes that the centrifugal force imbalance caused by rotation drives the circulation. Theoretical studies have demonstrated that large-scale meridional flows can exist in the radiative zones of rotating stars \citep[e.g., see][]{2000stro.book.....T, 2009pfer.book.....M}. Meridional flows primarily facilitate the redistribution of angular momentum \citep[e.g.][]{1992A&A...265..115Z} and the mixing of elements \citep[e.g.][]{1979ApJ...229..624S}. They can transport angular momentum or elements in the form of a closed loop \citep{2009A&A...495..271D, 1983ApJ...267..334T, 2002A&A...390..561M}, and thus can be termed meridional circulation. The effect of the mixing process within stellar interiors on surface elemental abundances manifests as an increase or decrease. For low-mass main sequence (MS) stars, the depletion of surface Li is more likely to occur. Meridional circulation in the radiative zone of rotating stars can disrupt the boundary between the surface convective zone and the radiative zone, bringing Li into the radiative zone and thereby driving Li depletion \citep{1988ApJ...334..746C, 1989ApJ...347..821C}. Recently, \cite{2025ApJ...982L...4L} (hereafter Paper I) has provided the diffusion coefficient for radial mixing due to meridional circulation, derived from the radial mass changes in giants during material transport driven by meridional circulation. This is different from the view of \citet{1988ApJ...334..746C}. Based on asteroseismological evidence, the radiative zone of a rotating low-mass MS star can usually be regarded as maintaining nearly uniform rotation \citep[e.g.][]{2002RvMP...74.1073C, 2015MNRAS.452.2654B, 2016A&A...593A.120V, 2022A&A...662A..58V, 2020MNRAS.491.3586L, 2020MNRAS.491..690J}. This implies that the work of Paper I can be extended to low-mass MS stars. Consequently, in this paper, we investigate the role of meridional circulation on the Li anomaly behavior of Population I stars during the MS, based on the model of Paper I. In this study, we focus on two primary Li depletion issues during the MS. The first is the solar Li problem and the Li evolution of solar analogs. The second is the Li dip phenomenon observed in some open clusters.

A concise description of the radial mixing due to meridional circulation and the model setup is provided in Section \ref{sect2}. Sections \ref{sect3} and \ref{sect4} present the comparison of our model results with observations in two aspects: Section \ref{sect3} examines Li depletion in solar analogs, while Section \ref{sect4} covers the Li dip. The discussion and summary are given in Section \ref{sect5}.

\section{Method}\label{sect2}
\subsection{Radial Mixing for Meridional Circulation} \label{sect21}
In Paper I, we provided the diffusion coefficient for radial mixing due to meridional circulation, derived from the change in matter during material transport
\begin{equation}\label{eq1}
	D_{\rm MC} = 2\, \pi\, V_{\rm MC}\, \frac{r^2}{\Delta r},
\end{equation}
where $D_{\rm MC}$, $r$, and $V_{\rm MC}$ are the diffusion coefficient, radial position, and radial velocity, respectively. 
Drawing from \cite{1979ApJ...229..624S}, Paper I primarily examined the transport of products from the hydrogen-burning shell to the surface of giants. In contrast, for the MS stars, the direct impact of meridional circulation is the migration of material from the star's surface to the high-temperature regions within the star \citep[e.g.][]{1988ApJ...334..746C, 1992A&A...255..191C}. Thus, following the approach outlined in Paper I, we can derive
\begin{equation}\label{eq2}
	\overline{D_{\rm MC}} = 2\, \pi\,  \overline{V_{\rm MC}}\,\frac{r_{\rm 2}^2}{r_{\rm 1}-r_{\rm 2}}.
\end{equation}
The overbar on the variable denotes the mean value. Subscripts 1 and 2 correspond to the two ends of the radial mixing region due to meridional circulation. Paper I showed that our work is only suitable for the uniformly rotating radiative zone, but not for the radiative zone with strong differential rotation or the stellar core \citep[see][]{1997ARA&A..35..557P}. The boundary determination of the mixing region should be strongly restricted. Specifically, \( r_1 \) represents the distance from the center of the star to the base of the surface convective zone, while \( r_2 \) is the inner boundary of the rotation-driven mixing region and is determined as follows:
\begin{equation}\label{eq3}
r_2= \left\{
\begin{array}{lll}
	r_{\rm i}\ \ (r_{\rm i} > r_{\rm ii}) \\
	r_{\rm ii}\ \  (r_{\rm i} \le r_{\rm ii}) \\
	r_{\rm iii}\ \  (r_{\rm i} \to 0\ \text{and}\ r_{\rm ii}\to 0 )
\end{array}
\right.
\end{equation}
Where $r_{\rm i}$, determined by Equation (16) in Paper I, is the inner boundary of the uniform rotation zone. $r_{\rm ii}$ is the upper boundary of the core convective zone of the star. The convective instability condition is $\nabla>\nabla_{\rm ad}$\footnote{Temperature gradient, where $\nabla=\frac{d\,\text{ln}\,T}{d\,\text{ln}\,P}$ and $\nabla_{\rm ad}=(\frac{\partial\,\text{ln}\,T}{\partial\,\text{ln}\,P})_{\rm ad}$.}, and the boundary of the convective zone is determined by the Schwarzschild criterion \footnote{Here, we use MESA's default boundary placement algorithm, which determines the boundaries by evaluating the sign of $\nabla-\nabla_{\rm ad}$ in the discrete grid. While this method occasionally causes convergence failures in certain models \citep[e.g.][]{2018ApJS..234...34P}, we encounter no such issues during the current model operation. Notably,  MESA provides two schemes to improve this problem, i.e., predictive mixing and  convective premixing.}. $r_{\rm iii}$ is the center of the star. If the entire interior of the star satisfies the rigid body rotation criterion, then denote as $r_{\rm i}\to 0$. Similarly, if the star has no core convective zone, then $r_{\rm ii}\to 0$.

According to Paper I, the specific expression of diffusion coefficient is obtained
\begin{equation}\label{eq4}
	\overline{D_{\rm MC}} = 2\,\pi\, \frac{r_2^2}{r_1-r_2}\, \frac{R}{\tau_{\rm MC}} \ \text{with}\ \tau_{\rm MC} = \frac{G^2\,M^3}{3\,L\,\Omega_0^2\, R^4}.
\end{equation}
Here, $R$, $M$, $L$, and $\Omega_0$ denote the stellar radius, mass, luminosity, and surface angular velocity of the star, respectively. $G$ represents the gravitational constant, and $\tau_{\rm MC}$ is the timescale of meridional circulation. For low-mass MS stars, $\tau_{\rm MC}$ is $\sim 10^{12}\,\rm yr$, indicating that meridional circulation is active throughout the MS phase of their evolution.

\subsection{Modeling}\label{sect22}	
We explore the Li depletion behavior of MS stars by constructing stellar models. The reader can consult Paper I for a more detailed mathematical description of the modeling. We employ the Modules for Experiments in Stellar Astrophysics (MESA; version 11701) for modeling, with fundamental settings outlined in Paper I. We provide a representative MESA case with inputs of $1.0\,M_\odot$ and $Z=0.02$.\footnote{The configuration and output files of our MESA model are openly accessible on Zenodo (\url{https://zenodo.org/records/15787854}).} For decisions regarding initial mass, initial metallicity, input Li abundance, and input rotation velocity, see Sections \ref{sect31} and \ref{sect41}. Our models incorporate a single non-standard process: meridional circulation.

\section{Li Depletion for Solar Analogs} \label{sect3}
\subsection{Model Inputs}\label{sect31}
We gather samples of solar analogs from \citet{2019MNRAS.485.4052C}, \citet{2023MNRAS.525.4642R}, and \citet{2024arXiv240810999R}, as depicted in Figure \ref{f1}. The distribution of Li abundance in stars near the zero-age main sequence (ZAMS) is relatively broad, yet it predominantly clusters around \(2.3\,\text{dex}\). It is less than the meteorite abundance \citep[$\rm 3.3\,dex$;][]{1998SSRv...85..161G} and the default initial value of the model ($A(\rm Li)$ is $\rm 3.4\,dex$ when $Z=0.02$). Consequently, our models cannot ignore the Li depletion in the pre-MS stage. Here, we mainly focus on solar analogs. Then we can take the abundance of meteorites as the input Li abundance, i.e., \(A(\text{Li})_{\text{ini}} = 3.3\,\text{dex}\)\footnote{The Li abundance here is the input value of the model (initial abundance), not the abundance value at the ZAMS.}.
Simulations by \citet{2014ApJ...780..159E} revealed a positive correlation between the rotation period of low-mass stars in the MS phase and their age. For a \(1.0\,M_{\odot}\) star, the rotation velocity is around \(10\,\text{km}\,\text{s}^{-1}\) in the early MS, and it may decrease to about \(1\,\text{km}\,\text{s}^{-1}\) in the late MS. The equatorial rotation velocity, $V_{\rm e}$, of the current Sun is roughly \(2\,\text{km}\,\text{s}^{-1}\). 
Figure \ref{a1} shows the evolution of stellar equatorial velocities, where initial velocities at the ZAMS $V_{\rm ZAMS}$ are set to 10 and $2\,\rm km\,s^{-1}$. The equatorial velocity $V_{\rm e}$ exhibits faint variation during the MS stage. In terms of observation, launching velocity is nowhere near velocity at 1 Gyr or even 0.1 Gyr; these can differ by up to a few orders of magnitude \citep[e.g.][]{2018AJ....155..196R}. 
We therefore employ multiple initial velocities in our models due to their limited capacity to accurately simulate this complex velocity evolution. For solar analogs, our model inputs are: \(1.0\,M_{\odot}\), \(Z = 0.02\) ([Fe/H] = \(0\,\text{dex}\)), \(A(\text{Li})_{\text{ini}} = 3.30\,\text{dex}\), and \(V_{\text{ZAMS}} = 10\,\text{km}\,\text{s}^{-1}\) (or \(2\,\text{km}\,\text{s}^{-1}\)).
Given the paper's focus on mixing processes, it is especially important to perform mixing length coefficient, \(\alpha_{\text{MLT}}\), calibrations. Following \citet{2019ApJ...881..103Z}'s methodology, we calibrate the model of solar parameters incorporating meridional circulation, with results detailed in Table \ref{t1}. The optimal calibration value of 1.78 is subsequently applied to all stellar models. 
More detailed descriptions of $\alpha_{\rm MLT}$ calibration in stellar models are provided in \citet{2018ApJ...856...10J} and \citet{2023Galax..11...75J}. All of our models evolve from the pre-MS to \(X_c = 0.01\), where \(X_c\) represents the mass fraction of central hydrogen.

\subsection{Li Abundance versus Stellar Age}\label{sect32}
\begin{figure*}[hbt]
	\centering
	\includegraphics[width=16cm]{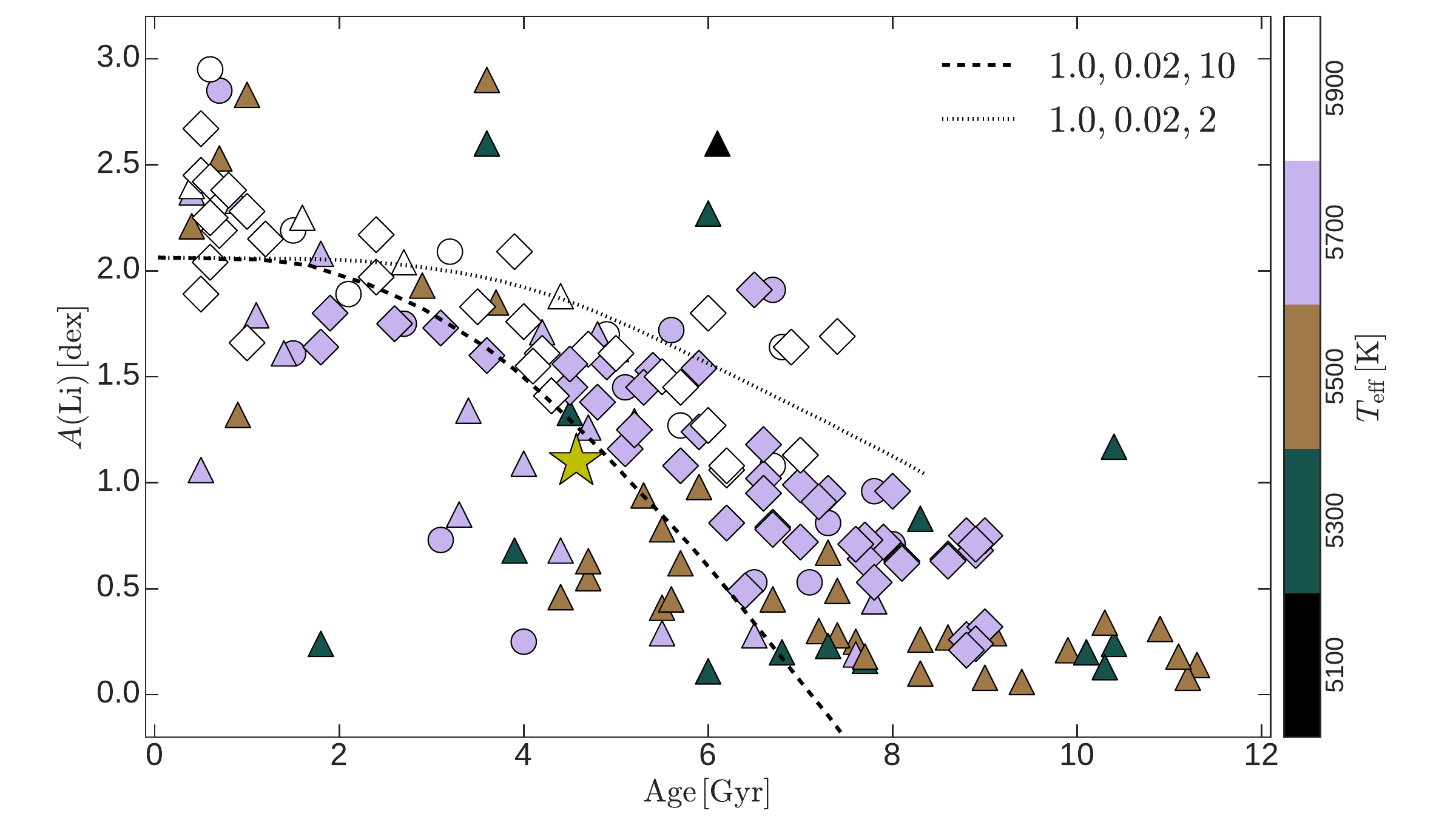}
	\caption{$A(\rm Li)$ vs. Age during the MS phase. The colorbar represents $T_{\rm eff}$. The scatter points in this figure are the samples of solar analogs: triangle \citep{2023MNRAS.525.4642R}, circle \citep{2024arXiv240810999R}, and diamond \citep{2019MNRAS.485.4052C}. The Sun is marked with a yellow star ($A(\rm  Li)\sim1.10\,dex$ \citep{1998SSRv...85..161G}, $\rm Age\sim 4.57\,\rm Gyr$, and $T_{\rm eff}\sim 5777\,\rm K$). The lines are the $A(\rm Li)-Age$ relationship predicted by the models ($1.0\,M_{\odot}$ and $Z=0.02$), with input velocities of 10 and $\rm 2\,km\,s^{-1}$, respectively.}
	\label{f1}%
\end{figure*}
\begin{figure*}[hbt]
	\centering
	\includegraphics[width=16cm]{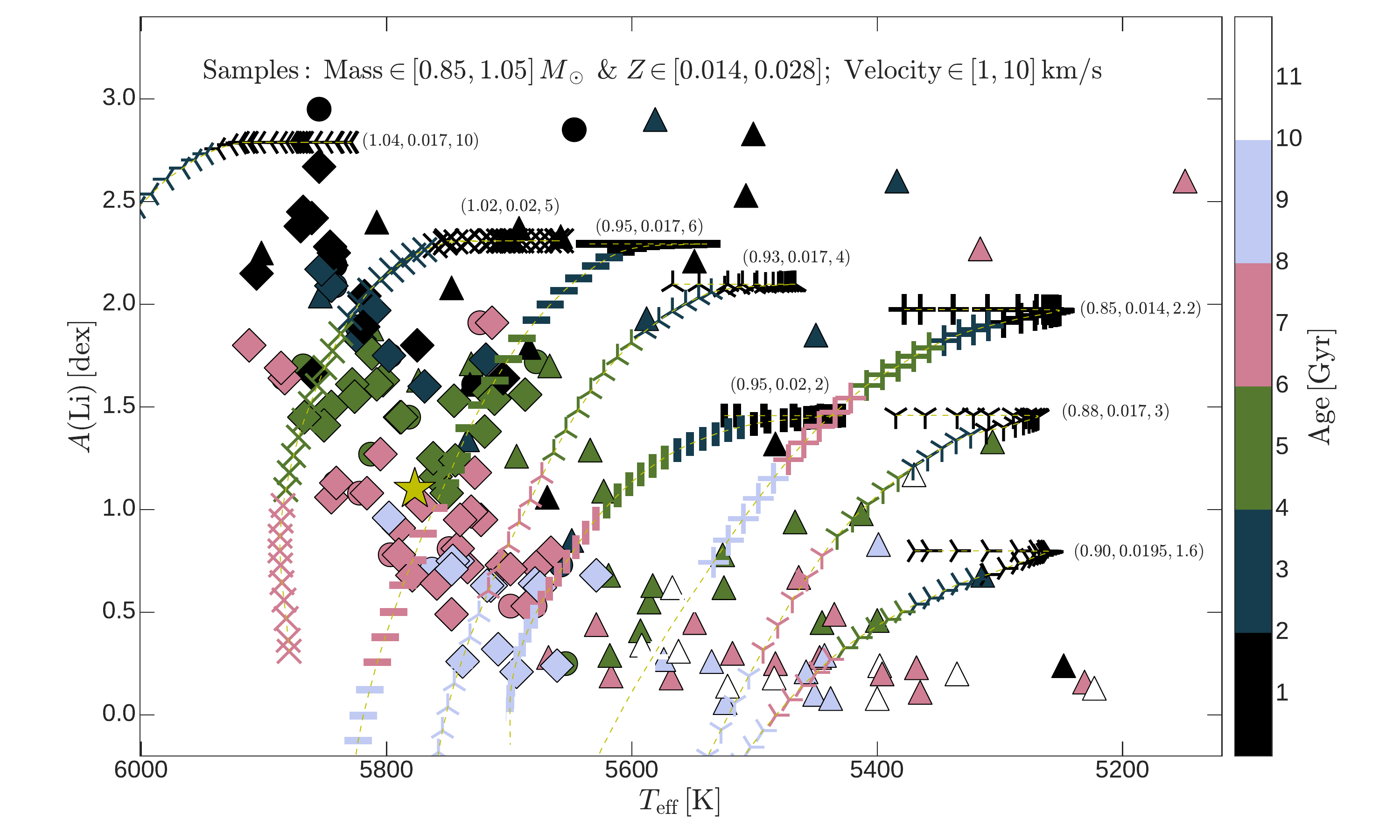}
	\caption{$A(\rm Li)$ vs. $T_{\rm eff}$. The colorbar represents stellar age. The meaning of the geometric symbols is consistent with Figure$\,\ref{f1}$. The gold dashed lines link the evolutionary trajectas of the star models, and the differences between the models are represented by different symbols. The input of the models are reduced to `(mass, metallicity, velocity)', and it corresponds to the evolutionary trajectory one by one. For the samples of solar analogs: $0.85\,M_{\odot}\le M \le 1.05 \,M_{\odot}$ and $0.014\le Z\le 0.028$, and the velocity range is referred to \citet{2014ApJ...780..159E}.}
	\label{f2}%
\end{figure*}
Figure \ref{f1} shows the \(A(\text{Li})\)-Age relationship of the observed samples along with the Li abundance evolution trajectories predicted by our models. Two evolutionary trajectories can cover most of the observed samples but are poorly matched to the Sun. Considering the evolution of the rotation period \citep[see][]{2014ApJ...780..159E}, for a star with \(1.0\,M_{\odot}\) and \(Z=0.02\), the actual evolution line can be expected to lie between the two trajectories and close to the dotted line. This is because the velocity of the star changes more dramatically in the early MS than in its late stage. For the Sun, before \(4.6\,\text{Gyr}\), the increment in velocity is around \(-8\,\text{km}\,\text{s}^{-1}\), and after that, the change in velocity is only \(\sim 1\,\text{km}\,\text{s}^{-1}\). Even so, our models still predict a higher solar Li abundance (approximately \(1.5\,\text{dex}\)). The \(A(\text{Li})\)-Age correlation of solar analogs can be fitted by linear relationships \citep{2019MNRAS.485.4052C, 2023MNRAS.522.3217M, 2023MNRAS.525.4642R}. \citet{2019MNRAS.485.4052C} pointed out that the Sun has actually experienced unusual Li depletion, since it is significantly out of the \(A(\text{Li})\)-Age correlation.

From the distribution of the solar analog samples, the degree of attenuation for Li abundance is low for ages less than \(\sim 3\,\text{Gyr}\), while the loss of Li in the subsequent time is about two orders of magnitude. Our rotation mixing models predict an increasing Li depletion behavior with age in solar analogs, which is well matched to observed samples.

\subsection{Li Abundance versus Effective Temperature}\label{sect33}
In addition, Figure \ref{f1} shows what appears to be an inverse correlation between stellar Li abundance and effective temperature. Overall, stars with higher \(T_{\rm eff}\) tend to have relatively higher \(A(\text{Li})\). This phenomenon has been observed in many open clusters \citep[e.g.][]{2021A&A...654A..46D, 2023ApJ...952...71S}. Because nuclear fusion reactions inside stars are ongoing, the energy generated causes the internal temperature and pressure to gradually increase, which in turn raises the surface temperature. Consequently, stellar effective temperature gradually increases during the MS phase. This suggests that simulating MS Li depletion based solely on the natural evolution of stars does not fully capture the evolution of real MS Li.

Since MS stars evolve toward higher \(T_{\rm eff}\), the Li abundance trend predicted by the models does not match observations. To address this, we can use isochrone information to correlate diverse stellar models with observed samples. We establish the \(A(\text{Li})\)-\(T_{\rm eff}\) relationship for these solar analogs in Figure \ref{f2}, with age as the third dimension. Across the entire sample of solar analogs we collected, the range of mass and metallicity is \(0.85\,M_{\odot} \leq M \leq 1.05\,M_{\odot}\) and \(0.014 \leq Z \leq 0.028\) (\(-0.183 \leq \text{[Fe/H]} \leq +0.164\)), respectively. For this class of stars, referring to \citet{2014ApJ...780..159E}, the rotation velocity distribution throughout the MS phase is around \(1-10\,\text{km}\,\text{s}^{-1}\). Based on the mass and metallicity information provided by the observations, we construct various stellar models. Moreover, low-velocity rotating stars are more common on the timescale of the MS, so we prefer to input a lower \(V_{\text{ZAMS}}\). The input information of the models is represented by `(\(M\), \(Z\), \(V_{\text{ZAMS}}\))' in Figure \ref{f2}. We use different colors to represent age, with observational samples shown geometrically and model results symbolically.

Figure \ref{f2} shows that for stars less than \(10\,\text{Gyr}\) old, our models can reproduce the distribution characteristics of Li abundance for observed samples. Due to the limited parameter range of the observed samples, our model results fail to reproduce the distribution law of Li abundance for stars older than \(10\,\text{Gyr}\). Since the older stars in Figure \ref{f2} are distributed on the low \(T_{\rm eff}\) side, which requires star models to have lower masses. Our models predict lower Li abundances for such stars, since they typically have stronger Li depletion due to a thicker surface convective zone \citep[e.g.][]{1965ApJ...142..174W, 1997ARA&A..35..557P}.

\section{Li Dip for Hyades and Praesepe} \label{sect4}
\subsection{Model Inputs}\label{sect41}

Unlike solar analogs, members of open clusters can often be considered as sharing cluster parameters, such as metallicity and age. Consequently, we will use the cluster's metallicity as an input for the stellar model. Moreover, assuming that the cluster members have similar ages, we will perform an isochronal fit to the model results. About mass, we treat it as a variable input. The relationship between \(A(\text{Li})\) and \(V\,\sin i\) suggests that the Li dip is related to rotation \citep[e.g.][]{1987PASP...99.1067B, 2017AJ....153..128C, 2023ApJ...952...71S}. Earlier work was limited by the velocity data, so the synergies of multiple physical processes were considered \citep[e.g.][]{1988ApJ...334..746C, 1993ApJ...416..312R}. Since we are primarily investigating the Li dip phenomena in the Hyades and Praesepe, we have collected more precise velocity samples, \(V\,\sin i\), of the members of the two open clusters \citep[see][]{2017AJ....153..128C}, and use them as a reference for the input equatorial velocity. The size of MS stars is relatively constant; therefore, the star models simulated by MESA will have a relatively constant rotation velocity (see e.g. Figure \ref{a1}). On the other hand, the rotation rate of low-mass stars slows down over time \citep{2007ApJ...669.1167B}. Since we are mainly concerned with the younger Hyades and Praesepe, we can assume that the input velocity, \(V_{\text{ZAMS}}\), is equal to the current stellar velocity. Referring to the meteoritic abundance \citep{1998SSRv...85..161G}, we assign an input Li abundance of \(3.3\,\text{dex}\) to all stellar models, i.e., \(A(\text{Li})_{\text{ini}} = 3.3\,\text{dex}\). On the other hand, the metallicity of the Hyades and Praesepe is higher than $Z_\odot$. For reference, we will allocate the initial Li abundance according to the default abundance composition of MESA, i.e., \(A(\text{Li})_{\text{ini}} = \rm [Fe/H]+3.40\,\text{dex}\) \citep[see][]{2024MNRAS.529.1423L}.
The main properties of the Hyades and Praesepe are as follows:
\begin{enumerate}
\item Hyades: Age is $0.72\,\rm Gyr$ and [Fe/H] is $+0.13\,\rm dex$ ($Z\sim0.0265$) \citep{2016A&A...585A.150N}. 
\item Praesepe: Age is $0.75\,\rm Gyr$ and [Fe/H] is $+0.16\,\rm dex$ ($Z\sim0.028$) \citep{2019A&A...623A.108B}.
\end{enumerate}
For each cluster, we adopt its metallicity as input to evolve stellar models to the cluster's age, recording $A(\rm Li)$, $T_{\rm eff}$, and $V_{\rm e}$ for stars reaching this evolution stage. These model outputs are plotted alongside observation data, and the model outputs are fitted. Here, the resulting curve is designated as the isochron.

\subsection{Results}\label{sect42}
\begin{figure*}[hbt]
	\centering
	\includegraphics[width=8.9cm]{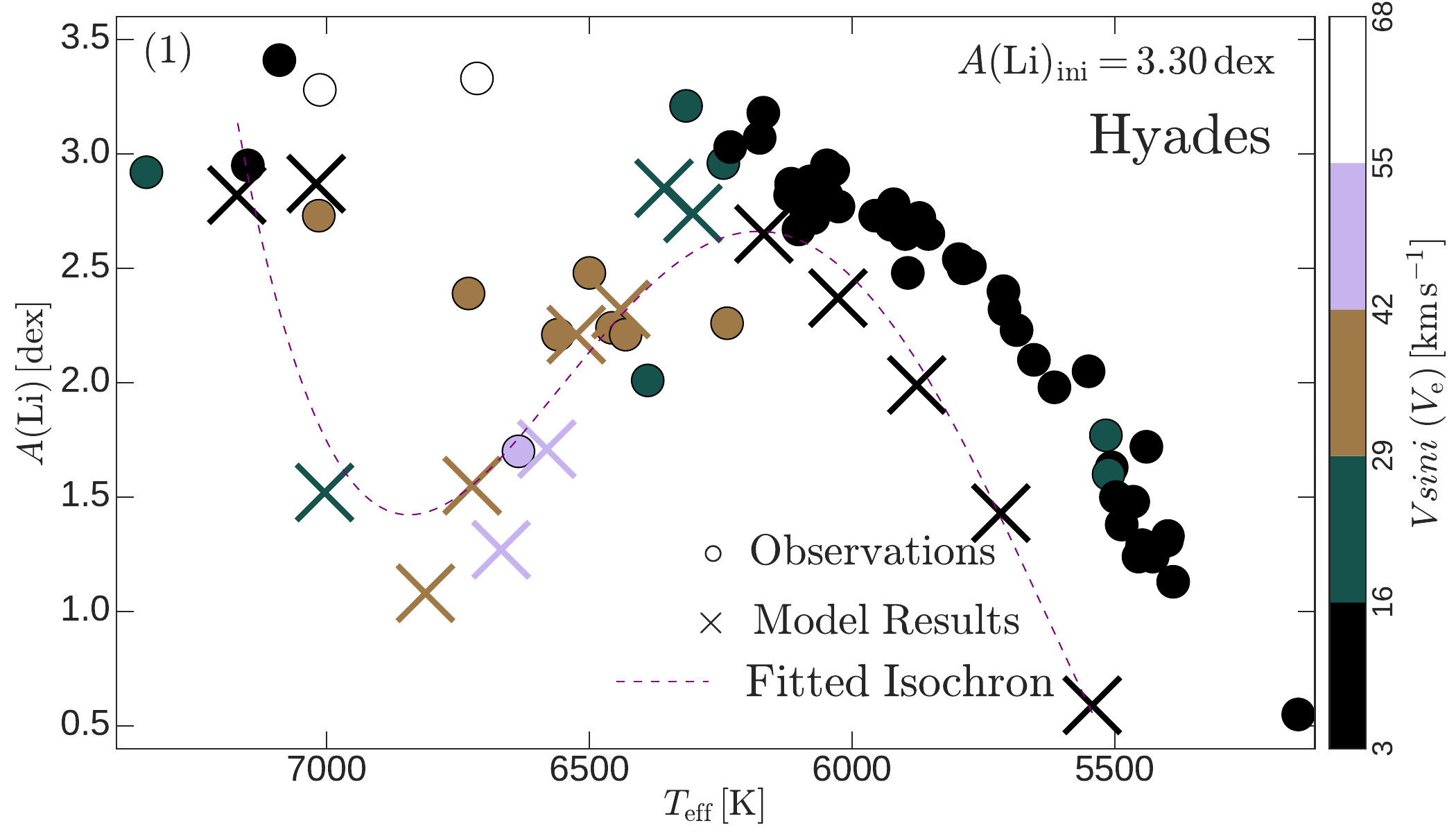}
	\includegraphics[width=8.9cm]{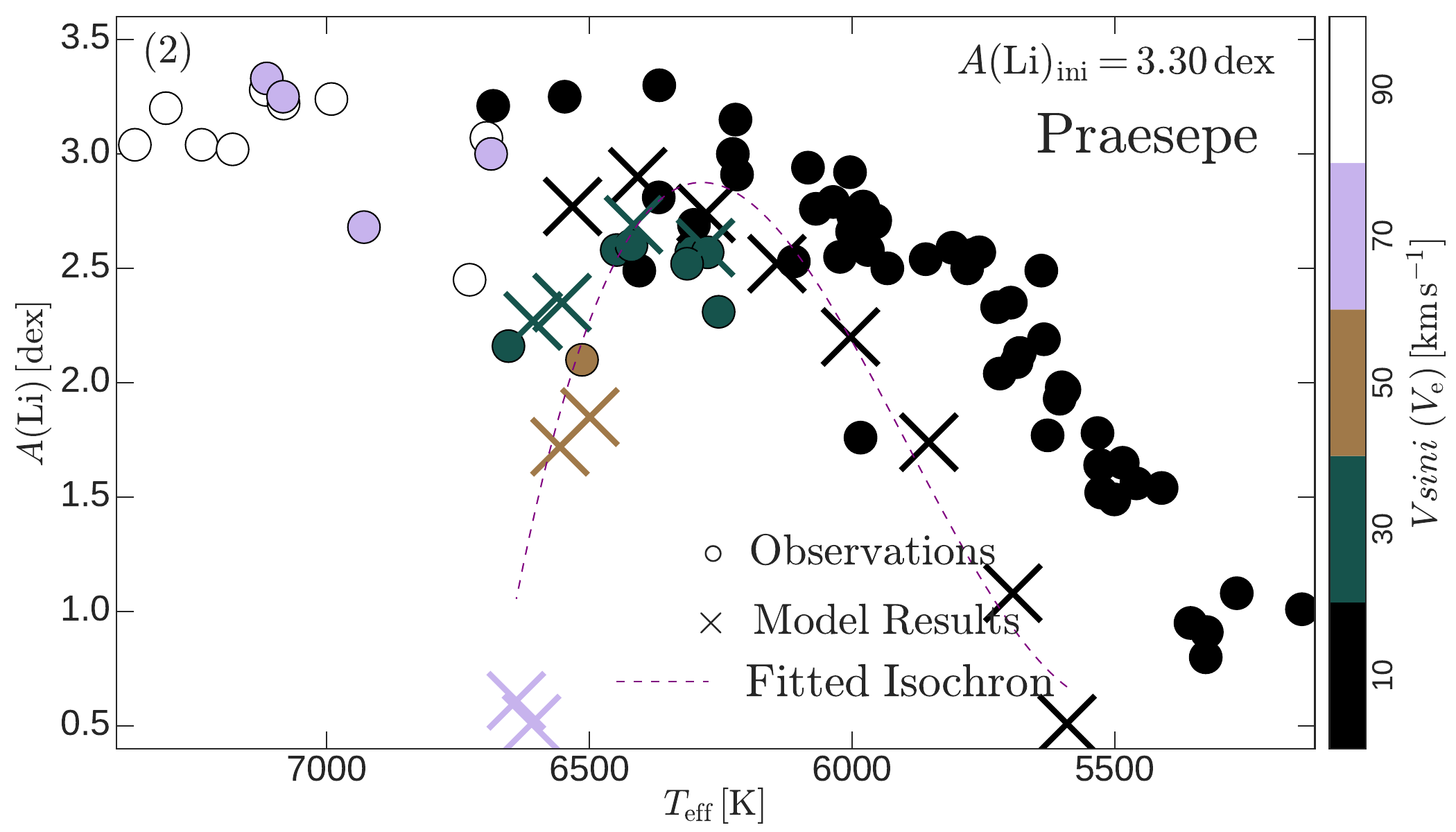}
	\includegraphics[width=8.9cm]{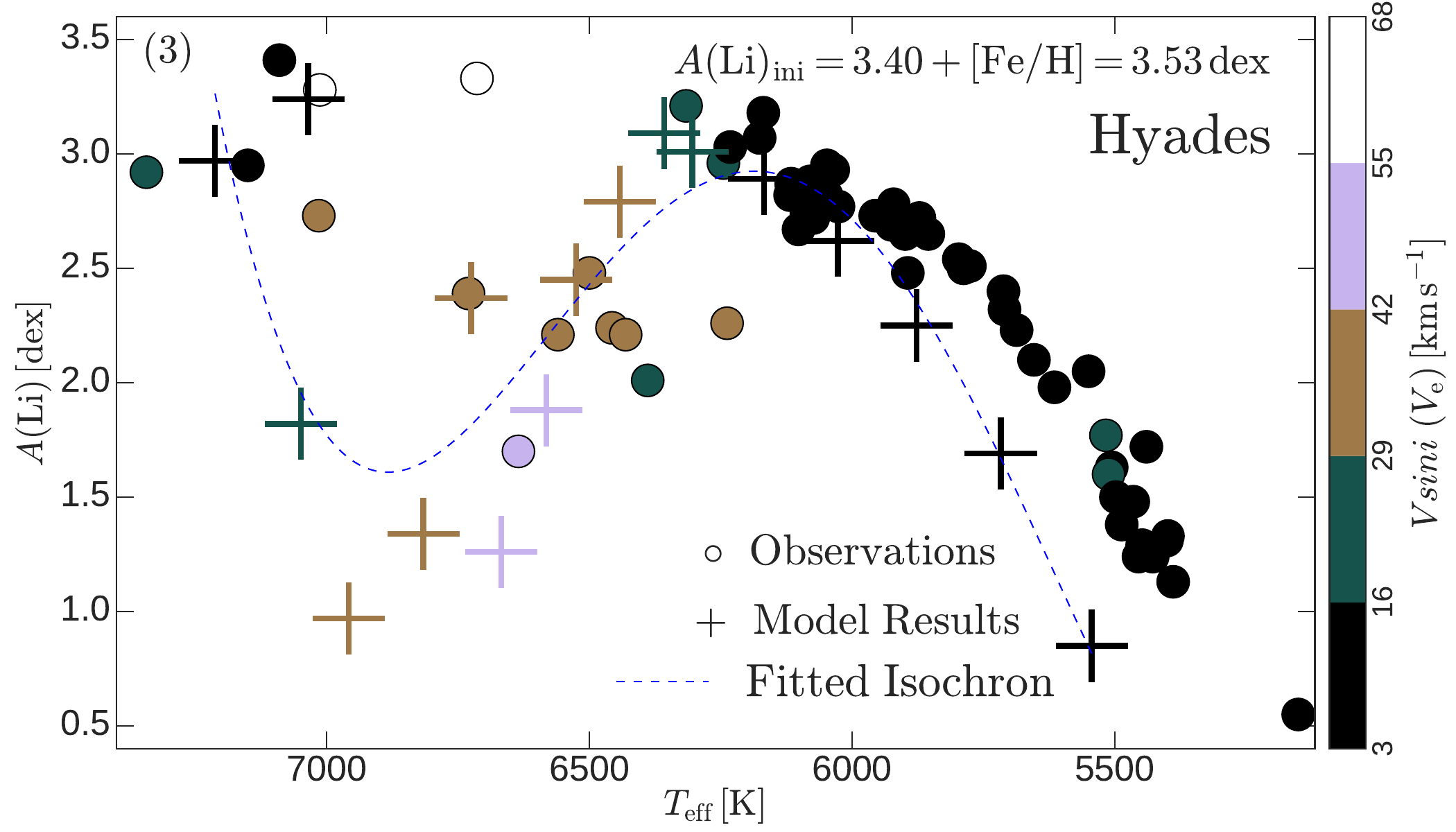}
	\includegraphics[width=8.9cm]{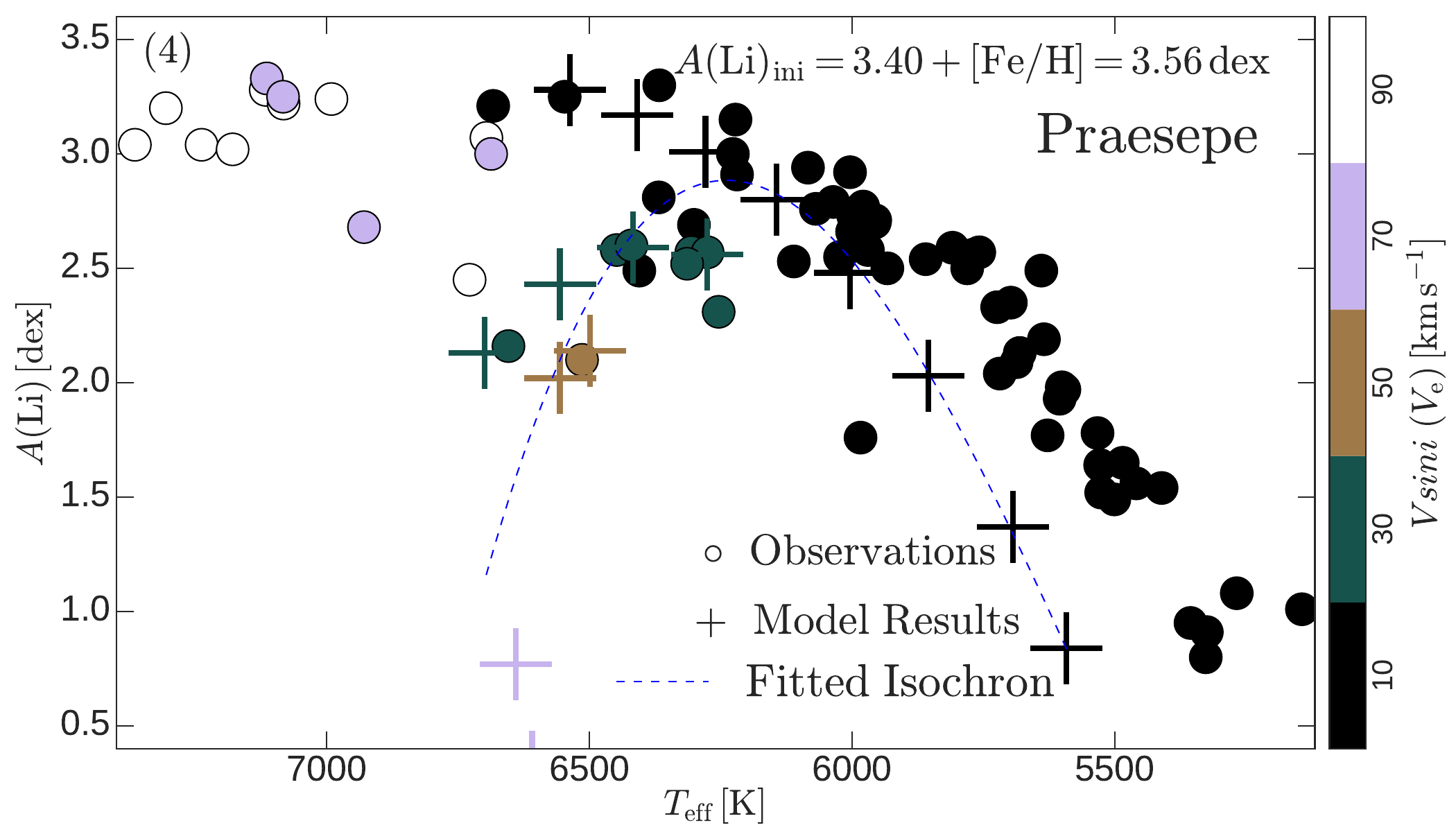}
	\caption{$A(\rm Li)$ vs. $T_{\rm eff}$. The samples of the Hyades and Praesepe are, respectively, from the Tables 6 and 7 of \citet{2017AJ....153..128C}. We mark the model results with `$\times$ (and +)' and fit them with dashed lines. The input Li abundance on the upper and lower panels is respectively $3.3$ and $\rm [Fe/H]+3.4\,dex$.}
	\label{f3}%
\end{figure*}
Our modeling results demonstrate distinct Li depletion patterns across stellar clusters (Figure~\ref{f3}). The isochrone fitting results reproduce the Li dip morphology in the Hyades and the distribution of Li abundance on the cold side of the Li dip in Praesepe (\(6600 > T_{\rm eff} > 6200\,\rm K\)). For the Hyades, sample stars within the effective temperature range of the Li dip exhibit relatively higher velocities, while those on both sides of the Li dip have lower velocities. As Equation~(\ref{eq4}) shows, \(\overline{D_{\rm MC}} \propto \Omega_0^2\) and \(V = \Omega_0 R\), implying that higher rotation velocities lead to more significant Li depletion. The velocity distribution of the samples in the Hyades presents a convex shape. Because of this feature, our models accurately recreate the Li dip in the parameter space of the Hyades members. For the Praesepe, however, the velocity distribution of the samples generally increases with \(T_{\rm eff}\), with the local velocity distribution showing the opposite trend only on the cold side of the Li dip. Therefore, while our models can fit the observations on the cold side of the Li dip in Praesepe, they do not match the Li distribution characteristics on the hot side (\(7000 > T_{\rm eff} > 6600\,\rm K\)).

Additionally, our models predict considerable change in Li abundance for stars with \(T_{\rm eff} < 6200\,\rm K\), resembling the observed downward trend in Li abundance. This trend stems from the behavior of Li depletion during the pre-MS stage. The evolution of Li abundance is strongly correlated with stellar parameters during both the MS and pre-MS phases. The Li abundance evolution during the MS is primarily constrained by the surface convective zone in the standard model. Higher low-mass stars with relatively thin surface convective zone maintain cooler convective zone, resulting in minimal impact on surface Li abundance. Conversely, lower low-mass stars, which require thicker convective zones to compensate for low internal temperature and pressure gradients, exhibit higher temperatures at the base of their surface convective zones, leading to more significant Li depletion. Moreover, metallicity plays a significant role in Li evolution; higher metallicities increase opacity, necessitating larger temperature gradients and allowing the convective zone to extend deeper into hotter regions. These factors influence Li evolution similarly during the pre-MS phase \citep[see][]{1997ARA&A..35..557P, 2017AJ....153..128C}. Within the effective temperature range of the Li dip, Li depletion is slight during both the pre-MS and MS phases. However, It becomes more pronounced with decreasing \(T_{\rm eff}\) at the cooler end of the Li dip. Our models reproduce distributions consistent with observation constraints when incorporating pre-MS Li depletion, except that the depletion amplitude exceeds observed levels.

\section{Discussion and Summary}\label{sect5}
This study investigates Li depletion mechanisms in Population I MS stars through the lens of meridional circulation-induced radial mixing. We focus on two pivotal phenomena: (1) the persistent Li depletion observed in solar analogs, and (2) the characteristic Li dip exhibited by mid-F type stars in open clusters. The following is the discussion and summary of the two parts.

\subsection{Li Abundance, Rotation Velocity, and Age}
Current observations demonstrate that rapidly rotating solar analogs generally exhibit higher Li abundances compared to their slower-rotating counterparts \citep{2010A&A...515A..93T, 2017A&A...602A..63B}. Our model predictions, however, appear inconsistent with this observational trend. To reconcile this discrepancy, we analyze the role of stellar age in Li evolution. The $A(\mathrm{Li})$-$\rm Period$ correlation in solar analogs mirrors their $A(\mathrm{Li})$-$\rm Age$ relationship, demonstrating an inverse dependence where $A(\rm Li)$ decreases with increasing rotation period. Combined with \citet{2014ApJ...780..159E}'s established Age-Period correlation for the Sun-like mass stars, we can catch a glimpse of self-consistent tripartite relationship: young stars with rapid rotation retain higher Li abundances, which subsequently decline as stellar spin down progresses during evolution. Figure~\ref{f1} reveals two key evolutionary phases: (1) During early MS stages, rotation velocity exerts minimal influence on Li depletion. Despite active rotation-driven mixing during this phase, rapid angular momentum loss limits the mixing duration, resulting in modest Li depletion \citep[e.g.][]{2016ApJ...829...32S}. (2) In later evolutionary stages, rotation-driven mixing weakens but produces cumulative depletion effects. While instantaneous depletion rates decrease, the accumulated Li depletion over time becomes significant. This temporal evolution of mixing efficiency explains why the observed $A(\rm Li)$-Rotation correlation emerges not from instantaneous velocity dependence, but rather from the coupled evolution of rotation and Li abundance over stellar lifetimes.
Thus, a positive correlation between the Li abundance of solar analogs and rotation velocity may be verified by our results and stellar evolution characteristics rather than single model results.

The morphology of the Li dip seems to be related to the cluster age. Observation evidence from \citet{2021FrASS...8....6R} reveals three evolutionary phases: (1) Young clusters (e.g., NGC 3532, \(\sim 0.4\,\text{Gyr}\)): The Li dip has appeared, but the depression is shallow, and its location (temperature range) may be adjusted with metallicity. For example, the Li dip in a high-metallicity cluster corresponds to higher-mass member stars \citep[tending toward the hot side, see e.g.,][]{2020A&A...640L...1R}. (2) Middle-aged clusters (e.g., NGC 2420, \(\sim 1.7\,\text{Gyr}\)): The depth and extent of the Li dip become more significant, and the depression may shift toward lower-mass (cooler) stars with age. (3) Extremely old clusters (e.g., Berkeley 39, \(\sim 6\,\text{Gyr}\)): The Li dip tends to stabilize, indicating that the mixing process is complete early in stellar evolution.
These evolutionary characteristics can be explained by the results presented in the previous paragraph. In the early stages, the rotation velocity is high. Although rotational mixing is strong, Li depletion has just begun, so the attenuation of Li is not significant. Soon afterward, the velocity decreases, and the total loss of Li becomes more pronounced, making the Li dip more distinct. In the later stages, the rotation velocity is very low, and Li is almost no longer depleted, so the Li dip stabilizes.
It is important to note that the depth and location of the Li dip are influenced by a variety of factors, including the temperature of the star (i.e., mass), metallicity, and the efficiency of the internal mixing process \citep[e.g.][]{2015A&A...576A..69D}. Therefore, the change in the depth of the Li dip over time is not a simple linear relationship but is influenced by a variety of complex factors. This means that the exact age dependence of the Li dip still needs to be verified with larger samples.

\subsection{Li Depletion}

The solar Li abundance predicted by our models is approximately three times higher than the Sun's current Li abundance, yet it aligns well with the Li distribution of solar analogs. This suggests that additional MS Li depletion processes, as discussed in Section \ref{sect1}, need to be incorporated into our models to account for the solar Li problem.
Given the multitude of physical processes that significantly contribute to Li depletion during the MS, accurately predicting solar Li abundance requires a nuanced consideration of the cumulative effects of these processes. Relying solely on a single physical process to explain Li abundance does not align with the complexity of the actual physical context.

Starting from the velocity distribution of low-mass stars in the MS phase, the evolution of our models (\(1.0\,M_\odot\) and \(Z=0.02\)) is consistent with the observed sample distribution. The effects of mass and metallicity are not considered here (see Figure \ref{f1}). However, when analyzing Li abundance and effective temperature, the optimal fit requires that both the mass and metallicity of our models be generally low, clustering around \(0.9\,M_{\odot}\) and \(Z=0.017\) (see Figure \ref{f2}). If we take \(Z=0.014\) as \(\rm [Fe/H]=0\,\rm dex\) \citep[see e.g.][]{2009ARA&A..47..481A}, the mass will increase slightly, but it will not concentrate at \(1.0\,M_{\odot}\). According to the standard model, a star with \(M>1.0\,M_{\odot}\) and solar metallicity does not experience significant Li depletion during the MS. Therefore, the observed Li distribution can only be attributed to lower-mass stars, and models with higher metallicities are expected to match samples with higher dispersion (mainly the diamond samples in Figure \ref{f2}; \(\sim 40\%\) high metallicity (\(\rm [Fe/H]>0\,\rm dex\)) samples). Overall, our models provide a reasonable explanation for Li depletion in solar analogs.

For lower mass stars in the pre-MS phase, our models predict excessive abundance decrease (see Figure \ref{f3}), highlighting limitations in current pre-MS modeling. This excessive Li depletion correlates strongly with metallicity. One model-based solution adopts the solar-calibrated value $Z_\odot = 0.014$. Furthermore, stellar models initialize metallicity using cluster values without accounting for intra-cluster individual differences. Several physical mechanisms offer promising mitigation pathways, including magnetic field effects \citep{2014MNRAS.445.4306J,2015ApJ...807..174S} and accretion processes \citep{2010A&A...521A..44B}.

\subsection{Li Dip}
\begin{figure*}[hbt]
	\centering
	\includegraphics[width=8.5cm]{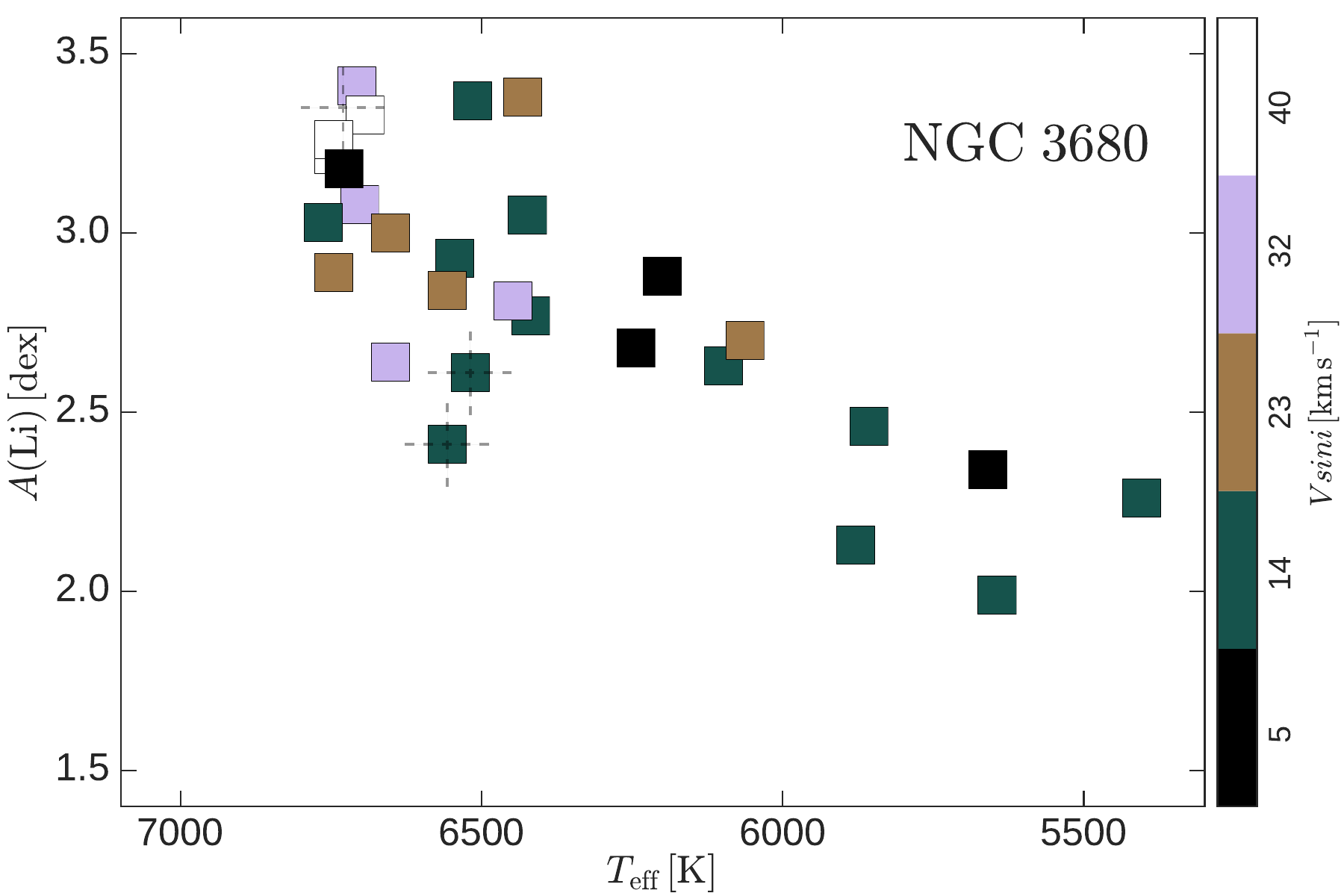}
	\includegraphics[width=8.5cm]{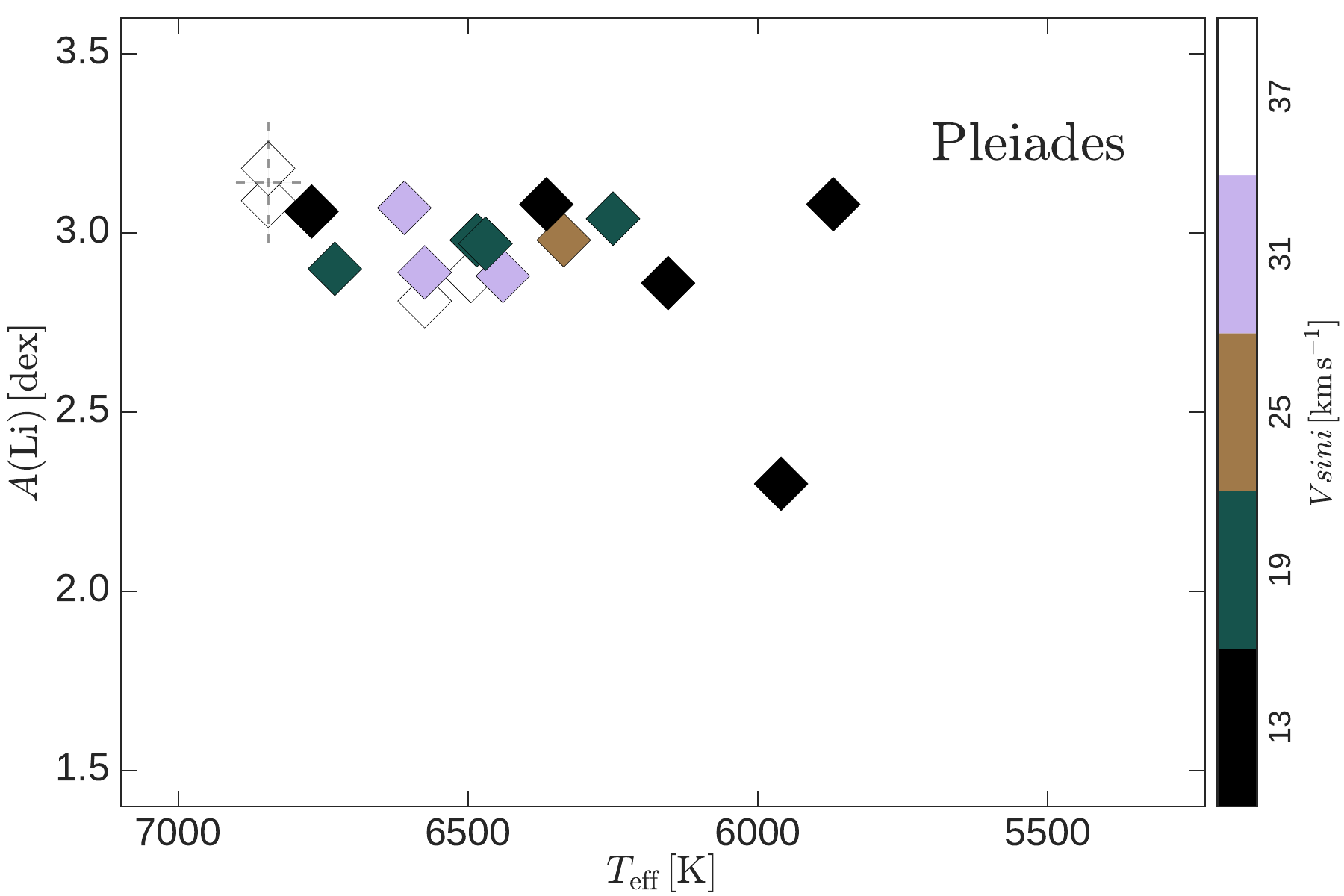}
	\caption{Similar to Figure \ref{f3}, but with sample stars from NGC 3680 \citep[$\sim\,\rm 1.8\,Gyr$,][]{2009AJ....138.1171A} and Pleiades \citep{1988ApJ...327..389B}. The sample stars, covered by the cross symbol composed of light gray dashed lines, signify that their distribution characteristics of Li abundances cannot be replicated by our models.}
	\label{f4}%
\end{figure*}
From Section~\ref{sect4}, it is evident that for our model to accurately simulate the Li dip behavior of a cluster, a convex distribution of rotation velocity in the corresponding temperature range is necessary. The velocity distribution of the sample stars in the Li dip region for the Hyades satisfies this condition. However, Praesepe does not fully meet this characteristic. Similar velocity distributions, either akin to or distinct from those of the two clusters, have also been observed in other Li dip clusters, such as the Pleiades \citep{1988ApJ...327..389B}, M48 \citep{2023ApJ...952...71S}, and NGC 3680 \citep{2009AJ....138.1171A}. The M48 samples of \citet{2023ApJ...952...71S} exhibit a clearly convex velocity distribution in the Li dip region, indicating a good match with our model results (see their Figure 2). Similar conditions are observed in Coma Berenices and Ursa Major \citep[see Figure 2 of][]{1988ApJ...334..746C}. However, a more complex situation arises in NGC 3680 (see Figure~\ref{f4}). Focusing on the samples marked by the cross symbol in Figure~\ref{f4}, for the higher rotation velocity samples on the high-temperature side of the NGC 3680 Li dip, our model results will predict stronger Li depletion. Conversely, weaker Li depletion is anticipated for the lower rotation velocity samples on the cold side. Ignoring these samples yields an isochrone fitting line similar to that of the Hyades. For the Pleiades, with an age of \(\sim 0.09\,\rm Gyr\) and [Fe/H] \(\sim 0.0\,\rm dex\) \citep{2019A&A...623A.108B}, the degree of the Li dip is relatively weak. Referring to the models in red for the Hyades (\(V\sim 35\,\rm km\,s^{-1}\), \(T_{\rm eff}\sim 6600\,\rm K\)), the Li depletion is about \(0.8\,\rm dex\). Considering that the Pleiades has lower age and metallicity than the Hyades, the Li depletion of the corresponding samples would be less than \(0.8\,\rm dex\). This implies that our models can account for the weak Li dip behavior in some clusters, contingent upon the velocity distribution.

Because the hot side region of Li dip is mixed with some high rotation velocity samples for most clusters, our models generally cannot fully explain the Li distribution of all sample stars in this region. Additional processes should be considered. For instance, \citet{1988ApJ...334..746C} proposed that radiative acceleration may inhibit Li depletion in the high-velocity samples on the hot Li dip side. Unlike the hot side, the velocity on the cold side of the Li dip in these clusters generally correlates with changes in \(T_{\rm eff}\), which can be characterized by the Li abundance distribution predicted by our models. In summary, our models can reproduce the Li abundance distribution on the cold side of the Li dip in open clusters but can only explain the hot Li dip side in some clusters.

It is important to note that the above discussion is based on \( V\,sin i \), with the input of $V$ adopted as $V\,sini$, while the real rotation velocity is \( V \). This is because it is not possible to accurately define \( sin i \) observationally. The Li dip regions that cannot be explained by our single models may also be affected by this effect. For example, low \( V\,sin i \) samples appear in the central region of the NGC 3680 Li dip. In addition, the modeling results are limited by the assumptions of uniform age and metallicity. As a result, we do not account for individual differences, which may be a key factor in addressing the model's limitations.

\begin{acknowledgments}

This research is funded by a grant from the National Basic Science Center Project of China (grant No. 12288102). The work is supported by National Natural Science Foundation of China (grant Nos: 12503043, 11973079, 12133011, 11973052, 12022304, 12090040, 12090044, 12173080, 12273104, and 12373036), the National Key R\&D Program of China Nos. 2021YFA1600400, 2021YFA1600402, the Natural Science Foundation of Yunnan Province (grant No. 202201AT070158), and the Yunnan Fundamental Research Projects (grant No. 202401AS070045). H.-L. Y. acknowledges support from the Youth  Innovation Promotion Association of the CAS and the NAOC Nebula Talents Program. J.-H. Z. acknowledges support from NSFC grant No. 12103063 and from China Postdoctoral Science Foundation funded project (grant No. 2020M680672).
\end{acknowledgments}




\appendix
\setcounter{figure}{0}
\setcounter{table}{0}
\section{Supplementary Visual Materials}

\begin{figure}[hbt]
	\centering
	\figurenum{A1}
	\includegraphics[width=8.8cm]{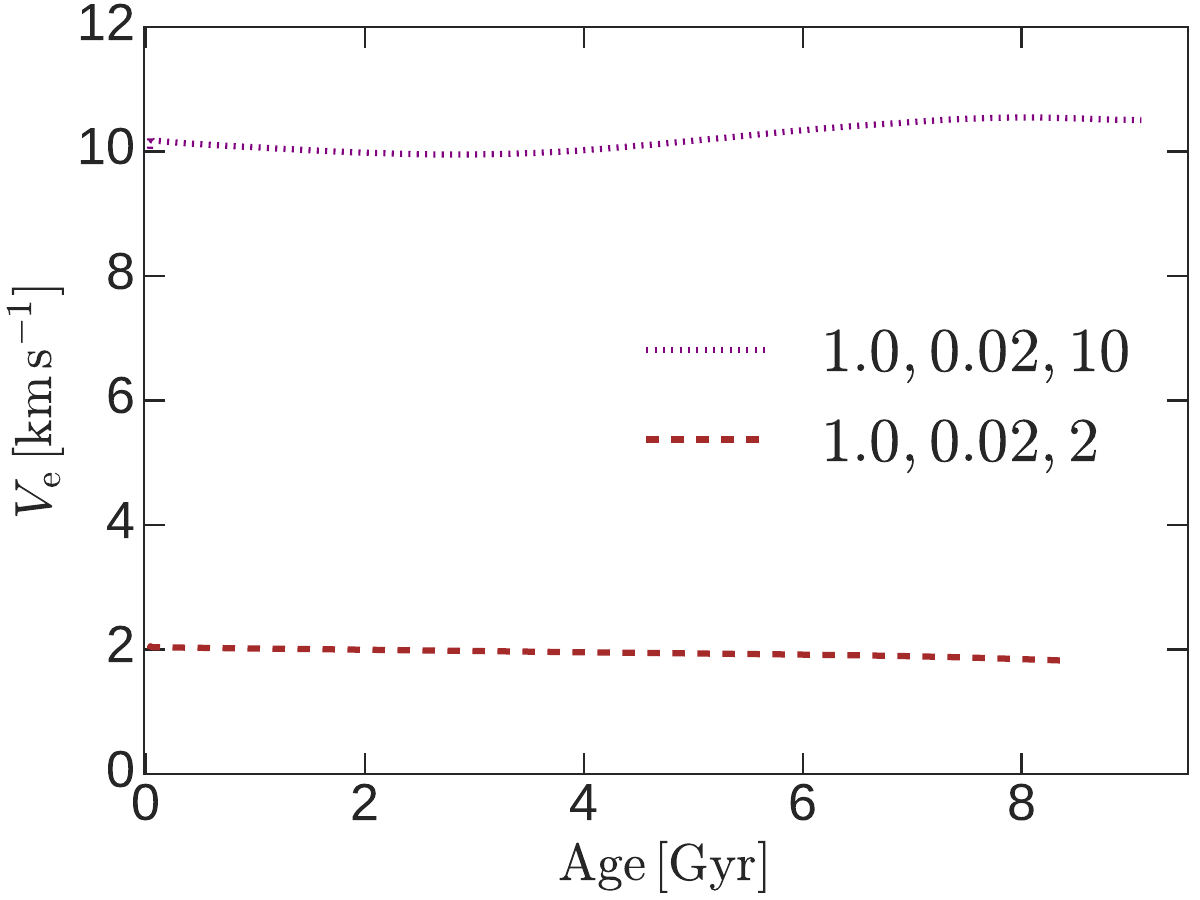}
	\caption{Equatorial velocity, $V_{\rm e}$, as a function of stellar age. The model mass and metallicity are $1.0\,M_\odot$ and 0.02 respectively, and the input velocity at the ZAMS is 2 and $10\,\rm km\,s^{-1}$.}
	\label{a1}%
\end{figure}

Figure \ref{a1} presents the evolutionary trajectory of the equatorial velocity of a star during the MS stage.

Table \ref{t1} shows the calibration results of our model's mixing length coefficient. The mixing length coefficient is the only input variable. We run all the models to the current age of the Sun and record the effective temperature, radius, luminosity, and  rotation velocity at this moment. The input parameters are $1.0\,M_\odot$ and $Z=0.02$, respectively.

\begin{table*}
	\caption{Calibration of the Mixing Length Coefficient.} 
	\label{t1}
	\centering          
	\begin{tabular}{lrrrrrrrrrrrr} 
		\hline \hline \noalign{\smallskip}
 $\alpha_{\rm MLT}$   &   2.20 & 2.10 & 2.00 & 1.95 & 1.90 & 1.85 & 1.80 &1.78& 1.75 & 1.70 & 1.65 & 1.60 \\
		\hline \noalign{\smallskip}
Age [Gyr]& 4.570 & 4.570 &  4.570 & 4.570 & 4.570 & 4.570 & 4.570 & 4.570 & 4.570 & 4.570 & 4.570&4.570\\
$T_{\rm eff}\ [\rm K]$ & 5928& 5896&  5861&  5843 & 5824& 5805& 5785&5776& 5763& 5742&5720&5696\\
$\rm log$$(R/R_\odot)$  &  -0.024&   -0.020&  -0.015&  -0.013 & -0.011& -0.008& -0.006&-0.004& -0.003& 0.000&0.003&0.006\\
$\rm log$$(L/L_\odot)$   &  -0.003&   -0.004&  -0.005&  -0.006& -0.007& -0.008& -0.008& -0.009&-0.009& -0.010&-0.011&-0.011\\
$V\rm _e\ [km\,s^{-1}]$  &1.944&1.944&1.943&1.942&1.941&1.943&1.943&1.943& 1.942&1.943&1.943&1.943\\
\noalign{\smallskip}
		\hline
	\end{tabular} 
\end{table*}

\FloatBarrier




\end{document}